\documentstyle{mn}

\title[Angular Correlation Function of $K^{\prime}\sim 19.5$ Galaxies and a Cluster at $z=0.775$]{The Angular Correlation Function of $K^{\prime}\sim 19.5$ Galaxies and the
Detection of a Cluster at $z=0.775$} 
\author[N. Roche, S. A. Eales, H. Hippelein and C. J. Willott]{Nathan 
Roche$^{1,4}$, Stephen A. Eales$^{1,5}$, Hans Hippelein$^{2,6}$
\newauthor
 and Chris J. Willott$^{3,7}$\\ 
$^1$Department of Physics and Astronomy,
      University of Wales Cardiff,
      P.O. Box 913,
      Cardiff CF2 3YB, Wales.\\
$^2$Max Planck Institut f\"ur Astronomie,
      K\"onigstuhl 17,
      69117 Heidelberg, Germany.\\
$^3$Department of Astrophysics,
      University of Oxford,
     Nuclear and Astrophysics Laboratory, Keble Road,
      Oxford OX1 3RH, England.\\
{$^4$ \verb"ndr@astro.cf.ac.uk"}\hspace{8mm}   
{$^5$ \verb"sae@astro.cf.ac.uk"}\hspace{8mm}
{$^6$ \verb"hippelei@mpia-hd.mpg.de"}\hspace{8mm}
{$^7$ \verb"c.willott1@physics.ox.ac.uk"}\hspace{8mm}
}

\input{psfig.sty}

\begin{document}

\maketitle
\begin{abstract}

We investigate  (i) the clustering environment of a sample of 5 radio galaxies at $0.7\leq z\leq 0.8$ and (ii) the galaxy angular correlation function, $\omega(\theta)$, on five $K^{\prime}$-band ($2.1\rm \mu m$) images covering a total  of 162.2 $\rm arcmin^2$ to a
completeness limit $K^{\prime}\simeq 19.5$, 

   Applying two methods --  counting galaxies within 1.5 arcmin of each radio galaxy, and using
a detection routine with a modelled cluster profile -- we detect
a cluster of estimated Abell richness  $N_A=85\pm 25$ (class 1 or 2), approximately centred on the  radio galaxy 5C6.75 at $z=0.775$. Of the other radio galaxies, two appear to be in less rich groups or structures, and two in field environments. The mean clustering environment of all 5 radio galaxies is estimated to be
of $N_A=29\pm 14$ richness, similar to that of radio galaxies at more moderate redshifts of $0.35<z<0.55$. 

The angular correlation function, $\omega(\theta)$, of the detected galaxies showed a positive signal out to at least $\theta\simeq 20$ arcsec, with a  $\sim 4\sigma$ detection of clustering 
for magnitude limits $K^{\prime}=18.5$--20.0.
The relatively high amplitude of $\omega(\theta)$ and its shallow scaling with magnitude limit  are
most consistent with a luminosity evolution model in which E/S0 galaxies are much more clustered than spirals ($r_0=8.4$ compared to 4.2 $h^{-1}$ Mpc)
and clustering is approximately stable ($\epsilon\sim 0$) to $z\sim 1.5$, possibly with an increase above the stable model in the clustering of red galaxies at the highest  ($z>1.5$) redshifts.

Our images also show a significant excess of close (1.5--5.0 arcsec separation) pairs
of galaxies compared to the expectation from $\omega(\theta)$ at larger separations.  
We estimate that a $11.0\pm 3.4$ per cent fraction of $K^{\prime}\leq 19.5$ galaxies are in close pairs in excess of the observed $\omega(\theta)$, if this is of the form  $\omega(\theta)\propto \theta^{-0.8}$. 
This can be explained if the local rate of galaxy mergers and 
interactions increases with redshift as  $\sim (1+z)^m$
with $m=1.33_{-0.51}^{+0.36}$.

\end{abstract}

\section{Introduction}

Statistical measures of the clustering of distant galaxies, and the identification of rich galaxy clusters at high redshifts, are both of great importance in  the study of galaxy evolution and the formation of structure in the Universe.
The clustering of  galaxies on the sky, as described in terms of the 
angular correlation function, $\omega(\theta)$, has been  
studied extensively at $\lambda\simeq 0.4$--0.9 $\mu \rm m$
wavelengths. The amplitude of $\omega(\theta)$ decreases on going faintward approximately as expected if galaxies undergo moderate luminosity evolution while their intrinsic
clustering remains approximately stable in proper co-ordinates 
(e.g. Roche et al. 1996; Roche and Eales 1998; Postman et al. 1998).

It is now possible to survey sufficiently large areas at near infra-red
wavelengths, e.g.
 the $2.2\mu \rm m$ $K$-band, to measure the galaxy $\omega(\theta)$ 
amplitude (Baugh et al. 1996; Carlberg et al. 
1997). Near-IR surveys, compared to those at visible light wavelengths, 
 contain a much larger proportion of early-type galaxies and are less sensitive to the effects of increased 
star-formation in the past, so may be particularly useful in studying the evolution of clustering separately from that of the star-formation rate.
Previously (Roche, Eales and Hippelein 1998, hereafter Paper I),
we measured the $\omega(\theta)$ amplitude to $K=20$ on 17 small
fields totalling 101.5 $\rm arcmin^2$, imaged with either the  Redeye near-infra-red camera on the Canada-France Hawaii Telescope or the Magic camera on Calar Alto. The $\omega(\theta)$ amplitude appeared to be relatively high compared to that from blue-band galaxy surveys of comparable depth (e.g. Roche et al. 1996), suggesting the intrinsic clustering of red E/S0 galaxies is  significantly stronger ($r_0\simeq  8$ $h^{-1}$ Mpc) 
than that of spirals
and that clustering is approximately stable 
($\epsilon\simeq 0$) with redshift, but more $K$-band data was needed to improve the statistics.

Some faint galaxy surveys have shown an excess in the number of close 
pairs of galaxies compared to the number expected from the inward extrapolation of $\omega(\theta)$ at larger scales (e.g. Infante, de Mello and 
Menanteau 1996). The excess pairs appear to be interacting or merging galaxies, and the number observed suggest some increase with redshift in the  merger/interaction rate, an important  finding as the merger rate evolution is of cosmological interest (Carlberg, Pritchet and Infante 1994). However,  there have been large differences between the close pair statistics of different datasets. For example, in Paper I, we found some excess of 2--3 arcsec separation pairs on the 12 fields
observed in good seeing with the Redeye camera, but not on the data from the Magic camera, which was of lower resolution due to both poorer seeing and a larger pixelsize. Similarly, 
our $R\leq 23.5$ Wide Field Camera survey (Roche and Eales 1998) found the $\omega(\theta)$ amplitude  at $2<\theta<5$ arcsec to be significantly
higher that at larger $\theta$, indicating an excess of close pairs, on
 one 1.01 $\rm deg^2$ field but not on a second 0.75 $\rm deg^2$ field, where again the seeing was poorer. These inconsistencies suggest that faint galaxy close pair counts are  critically dependent on data quality, and require similar analyses of further  deep
surveys. 
 
The identification of rich clusters of galaxies at high redshifts ($z>0.5$) is also of great interest -- their number, richness and other properties 
may provide important constraints on cosmological parameters (e.g.
Eke, Cole and Frenk 1996). At visible light wavelengths,
 the contrast of clusters against the field galaxies falls steeply with increasing cluster redshift, making detection difficult at $z>0.6$,
but this is in part due to the strong k-correction dimming of early-type galaxies. In the $K$-band, the k-correction produces a brightening to $z\sim 1$, for all galaxies, so clusters should be detectable to higher redshifts. Near-IR 
surveys
have already found rich clusters at redshifts as high as $z=1.27$
(Stanford et al. 1997) and possibly at $z=2.39$ (Waddington 1998).

Searches for high redshift clusters have often concentrated on 
areas containing high redshift QSOs or radio galaxies, as some of these sources do lie in rich
clusters and the AGN provides an
easily measured redshift. The clustering environment of radio galaxies is of interest in
itself for understanding their evolution. At low redshifts 
the more luminous (FRII) radio galaxies are generally found in field
environments, whereas  at $z\sim 0.5$, luminous radio galaxies are found in similar numbers in the field and in rich clusters, with the 
average environment being approximately that of an Abell class 0 cluster
(Yates, Miller and Peacock 1989; Hill and Lilly 1991).

 Yee and Ellingson (1993) explained this as an environmental influence on radio 
galaxy evolution, whereby sources in clusters  decrease much
 more rapidly in radio luminosity at $z\leq 0.5$ than those in the field. 
Wan and Daly (1996) found no observable differences between the 
powerful radio galaxies in field  and cluster environments, and concluded that the change in radiogalaxy environment with redshift is most likely due to an evolution of the clusters themselves  --
an increase in the typical intra-cluster medium pressure 
between $z\sim 0.5$ and $z\sim 0$ might surpress FRII activity in most 
rich clusters at lower redshifts and allow only
the less radioluminous FRI outbursts to occur. However,
to better understand this process, it is important to determine whether there is any further 
change in the mean radio galaxy environment at $z>0.5$.

In Paper I, 17 high redshift ($z_{mean}=1.1$) 6C radio galaxies
 were cross-correlated with the 
surrounding $K<20$ galaxies to estimate their mean clustering environment. 
No significant signal was detected in the cross-correlation function --  the upper limits were consistent with a mean radio galaxy environment similar to that at $z\sim 0.5$ (i.e. Abell 0 clusters) but
argued against a much richer environment (i.e. Abell 2 clusters). 
 However, at similar redshifts, some of the 3CR radio galaxies do inhabit rich clusters e.g.  3CR184 at $z=0.996$ (Deltorn et al. 1997)  and  3C336 at $z=0.927$ (Bower and Smail 1997).

In this paper we investigate the clustering environment of 5 radio galaxies 
at $0.7\leq z\leq 0.8$, observed in the  $K^{\prime}$-band with the large format
Omega camera on Calar Alto (see Section 2.1). The more moderate radiogalaxy redshifts and   larger field  areas compared to the Paper I survey should  improve our chances of finding clusters. Section 2 describes the observational data
and the detection of galaxies and stars, Section 3.1 the visual impression of the radio galaxy environments and  Section 3.2 the modelling of distant clusters. We search for clusters centred on the radio galaxies using two methods, described in
 Sections 3.3 and 4 
respectively. In Section 5  we investigate the galaxy $\omega(\theta)$, and in Section 6 the number of close
pairs of galaxies. Section 7 discusses the implications of all the results.
\section {Data}

\subsection {Observations}

Our dataset consists of images of 5 fields, each centred on a  radio galaxy
at $0.7\leq z\leq 0.8$ (Table 1). One source is from the 6C catalog of radio
galaxies with 151 MHz fluxes of at least 2.2 Jy, the other four from the 7C catalog (Willott et al. 1998, and in preparation) with a fainter flux
limit of 0.5 Jy (these have names beginning `5C', as they 
were first detected in a 5C survey). These 7C galaxies have radio luminosities
$L(151 \rm MHz)\simeq 10^{27.6}$ W$\rm Hz^{-1}$, much lower than 3C galaxies but
above the FRII/FRI divide.
 The five galaxies were selected from the catalogs solely on the basis of being in the desired redshift range, so should be an unbiased sample of
$0.7\leq z\leq 0.8$ radio galaxies.

The fields were observed in the $2.1\mu \rm m$ $K^{\prime}$-band (Wainscoat and Cowie 1992) on the nights of 13 and 14 December 1997, using the Omega 
 wide-field near-infrared camera at the prime focus of the 3.5m Calar Alto telescope (Sierra de los Filabres, Andalucia, Spain). This has a $1024\times1024$ pixel HgCdTe (1--2.5 $\mu \rm m$) array with pixelsize
0.3961 arcsec. Each field was exposed for 30 seconds 120 times, and the 120
exposured stacked together and flat-fielded during the observing run. For the purposes of flat-fielding, the
120 exposures were slightly offset in a grid pattern, reducing the areas of the final images (in which all pixels have a full 60 minutes exposure) to approximately
$870\times 870$ pixels or $5.75\times 5.75$ arcmin.
\begin{table}
\caption{Positions (equinox 2000.0), spectroscopic redshifts and $K^{\prime}$ magnitudes (as measured from this data) of the five radio galaxies}
\begin{tabular}{lcccc}
\hline
Galaxy    & R.A. & Dec. & z & $K^{\prime}$ \\
\smallskip
5C6.25  & 02:10:24.45 & +34:10:46.0 &  0.706 & 17.44\\           
5C6.29  & 02:11:05.85 & +32:56:44.2 &  0.720 & 16.61\\
5C6.43  & 02:12:02.69 & +34:02:18.1 &  0.775 & 17.95\\
5C6.75  & 02:14:01.20 & +30:26:14.3 &  0.775 & 17.05\\
6C0822  & 08:25:47.36 & +34:24:26.5 &  0.700 & 17.48\\
\hline
\end{tabular}
\end{table}

\subsection{Source detection}
 
    The Starlink {\sevensize PISA} (Position, Intensity and Shape Analysis) package, developed
 by M. Irwin, P. Draper and N. Eaton, was used to detect and catalog the
objects. The  chosen 
detection criterion was that a source must exceed an intensity threshold of
$1.5\sigma$ above the background noise, or approximately $21.0$ $K^{\prime}$ mag $\rm arcsec^{-2}$, in at least 5 connected
pixels. As in Paper I, {\sevensize PISA} was run with deblend (objects connected at the detection threshold may be split if they
are separate at a higher threshold) and the `total magnitudes' option.
There were a total of 3160 detections.

 Detections in four circular `holes' (radius 9 to 20 arcsec) around bright, saturated stellar images  were excluded in order to remove spurious noise detections, leaving a total 162.3 $\rm  arcmin^2$ area of usable data .
 Radial profiles were fitted 
to several non-saturated stars on each image using `pisafit', and from these profiles the average resolution on the reduced images was estimated as 
$FWHM=1.13\pm 0.06$ arcsec.
Star-galaxy separation was performed using plots of central against total intensity, normalized to the ratio from the fitted stellar profile. On these plots the stellar locus was separable from the galaxies to $K^{\prime}\sim 17$,
and a total
of 185 detections to this limit were classed as stars. All 
fainter objects were assumed to be galaxies, but
we later apply corrections to our results for the effects of faint star contamination. 

\subsection{Galaxy counts}

Figure 1 shows the differential number counts, in $\Delta(K^{\prime})=0.5$ mag bins, of objects classed as galaxies on these images, with field-to-field error bars. These are compared with galaxy counts from Paper I and a number of other
$K$-band surveys, plotted here assuming a mean colour $K^{\prime}=K+0.13$, and
non-evolving and pure luminosity
evolution (PLE) models, which assume $q_0=0.05$ and are the same as those of Roche and Eales (1998) but computed in the $K^{\prime}$-band rather than the $R$-band.

To  $K^{\prime}\simeq 19.5$, our galaxy counts agree well with those from previous surveys, including those using much deeper data, suggesting that our galaxy detection is virtually complete. Our counts  level out at $19.5<K^{\prime}<20.0$, where on the basis of the count gradient from models and deeper data (${d ({\rm log} N)\over dm} \simeq 0.30$) they are $\sim 32$ per cent incomplete,
 and fall off at $K^{\prime}>20$. 
The $K$ and $K^{\prime}$-band number counts are generally in good agreement with the PLE model over the entire $\sim 10$ mag range. 
\begin{figure}
\psfig{file=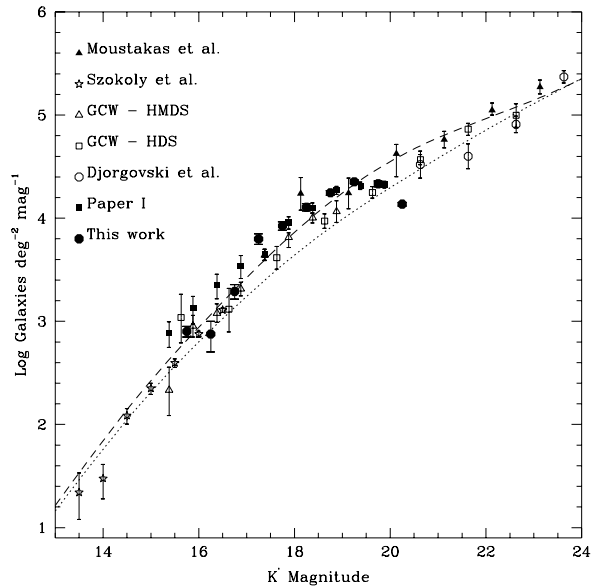,width=85mm}
\caption{Differential galaxy number counts, in 0.5 mag intervals of $K^{\prime}$
 magnitude, for our five fields compared to our Paper I and the  $K$-band surveys 
of  Gardner, Cowie and Wainscoat (1993), Djorgovski et al. (1995),
Moustakas et al. (1997) and Szokoly et al. (1998), together with the
 predictions of $q_0=0.05$ PLE (dashed) and non-evolving (dotted) models.} 
\end{figure}

Figure 2 shows the PLE model for the redshift distribution $N(z)$ 
at the $K^{\prime}=19.5$ limit. The predicted mean redshift is 0.94,  the radio galaxy redshifts are close to the peak of $N(z)$, and we expect to see galaxies to a maximum redshift $z\sim 2.4$, with E/S0s predominating at $z>1$.

\begin{figure}
\psfig{file=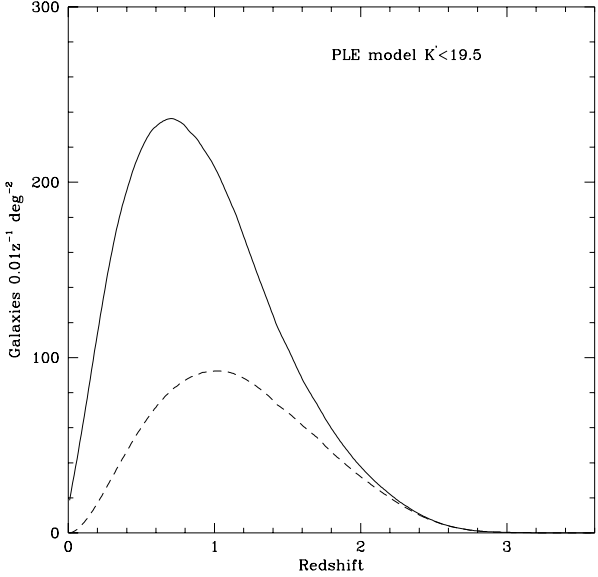,width=85mm}
\caption{Redshift distribution predicted by the PLE model for all galaxies (solid) and E/S0 galaxies only (dashed) to $K^{\prime}=19.5$.} 
\end{figure}

\subsection{Star counts}
Figure 3 shows the number counts of objects classed as stars, with field-to-field errors. The field-to-field scatter in star-counts is only $\sim \surd N$ as the 5 fields are at similar distances from the Galactic plane,
$|b|=25.9$--$33.5^{\degr}$.

To estimate the contamination of our galaxy sample by the unclassified $K>17$ stars, we fit the observed star counts with a model and extrapolate faintwards. 
To $K^{\prime}\sim 16.5$, the slope of the star counts is close to ${d({\rm log} N)\over dm}\simeq 0.2$, but $K$-band star-count models (Glazebrook et al. 1994; Minezaki et al. 1998)  flatten to ${d({\rm log} N)\over dm}\simeq 0.1$ at $17<K<20$. The  model of Minezaki et al. (1998) is for the Galactic pole and gives too shallow a
count at $14<K<17$ to fit our data, and the Glazebrook et al. (1994) 
models are only plotted to $K=17.8$, but the star-count models and observations of Minezaki et al. (1998) are quite similar to the  Reid et al. (1996) model for deep Galactic pole star-counts in the $I$-band, shifted by $I-K\sim 1.25$. 

We adopt the steeper Reid et al. (1996) $I$-band star-count model for the lower Galactic latitude of $b=45\degr$, shifted by $I-K=1.25$ and
extrapolated at $K<14.5$ with the Glazebrook et al. (1994) model for a $b=32\degr$ field. This is then normalized to our star-counts
at $12.5\leq K^{\prime}\leq 17.0$, with the best-fit normalization, corresponding to $1697\pm 356$ $\rm deg^{-2} mag^{-1}$ at $17.0<K^{\prime}<17.5$.
This model, shown on Figure 3, is then consistent with the observed counts within the statistical errors
(reduced $\chi^2=0.86$).

\begin{figure}
\psfig{file=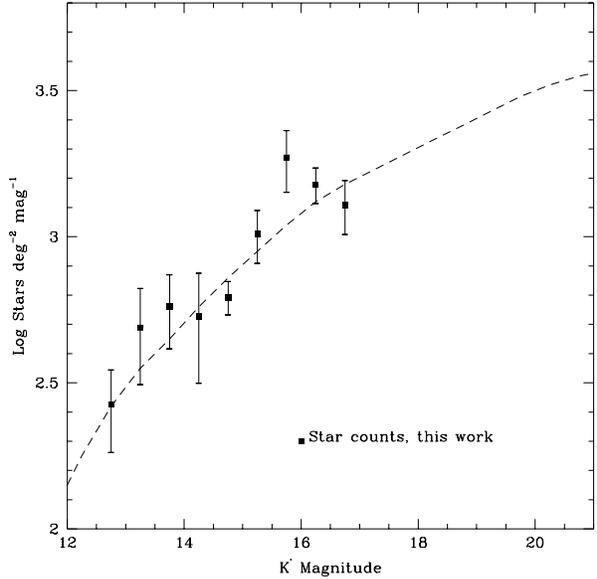,width=85mm}
\caption{ Differential star number counts in 0.5 mag intervals of $K^{\prime}$ magnitude, with field-to-field errors, and the adopted star-count model (dashed line).} 
\end{figure}

\section{Clustering around the radio galaxies}

\subsection{Visual impression} 

We first examine our data by eye 
for any obvious associations of galaxies around the radio
galaxies. Figure 4 shows maps of the distribution of $K^{\prime}\leq 19.5$ galaxies on each of 
the five fields, with the radio galaxies indicated. The images provide more
information than these maps as connecting filaments and interactions between galaxies may be visible. The visual impression is that 

(i) The radio galaxy 5C6.25 appears to lie on a long $S$-shaped filament, but
 as many of these galaxies are in filaments, this 
may not necessarily indicate an especially clustered environment. 

(ii) The radio galaxy 5C6.29 is not obviously in any sort of association
(the four bright galaxies nearby are almost certainly at much lower redshift).

(iii) 5C6.43 has a close companion of similar brightness ($K^{\prime}=18.01$), 6.4 arcsec away, with some signs of an interaction, but it is not obviously part of any larger group.

(iv) 5C6.75 lies within a more compact $S$-shaped structure, some 1 arcmin across with at least 10
bright members and a number of faint galaxies in the same area.
The region around the radio galaxy contains many interacting pairs and
small groups of galaxies.
The concentration of galaxies  is obvious on Figure 3 and seems a good candidate for being a true cluster.

(v) 6C0822 has two close companions of similar brightness and may be part of
an interacting group of several galaxies, although with fewer members than the association around 5C6.75.

Hence the radio galaxies appear to be in a wide range of environments, with 
evidence of clustering in some cases. To quantify this, we consider a model
rich cluster placed at the redshifts of the radio galaxies.

\subsection{Cluster modelling}

One measure of the richness of a galaxy cluster is the `Abell richness', 
$N_A$, defined as the number of cluster members with apparent magnitudes $\leq 2.0$ magnitudes fainter than the
third-ranked galaxy (e.g. Abell et al. 1989). Rich clusters are defined as those with $N_A\geq 30$, with $30\leq N_A\leq 49$ being Abell class 0, 
$50\leq N_A\leq 79$ class 1 and $80\leq N_A\leq 129$ class 2. 

The surface density of galaxies in rich clusters, in terms
of the projected distance from the centre $r=\theta d_A$, typically
follows a profile of the form
\begin{equation}
\rho(r)={\rho_0\over[1+(r/R_c)^2]^{\alpha}}
\end{equation}
where $R_c$ is the core radius,
out to $r_{max}$ where $\rho(r)$ falls to zero.

\onecolumn
\begin{figure}
\psfig{file=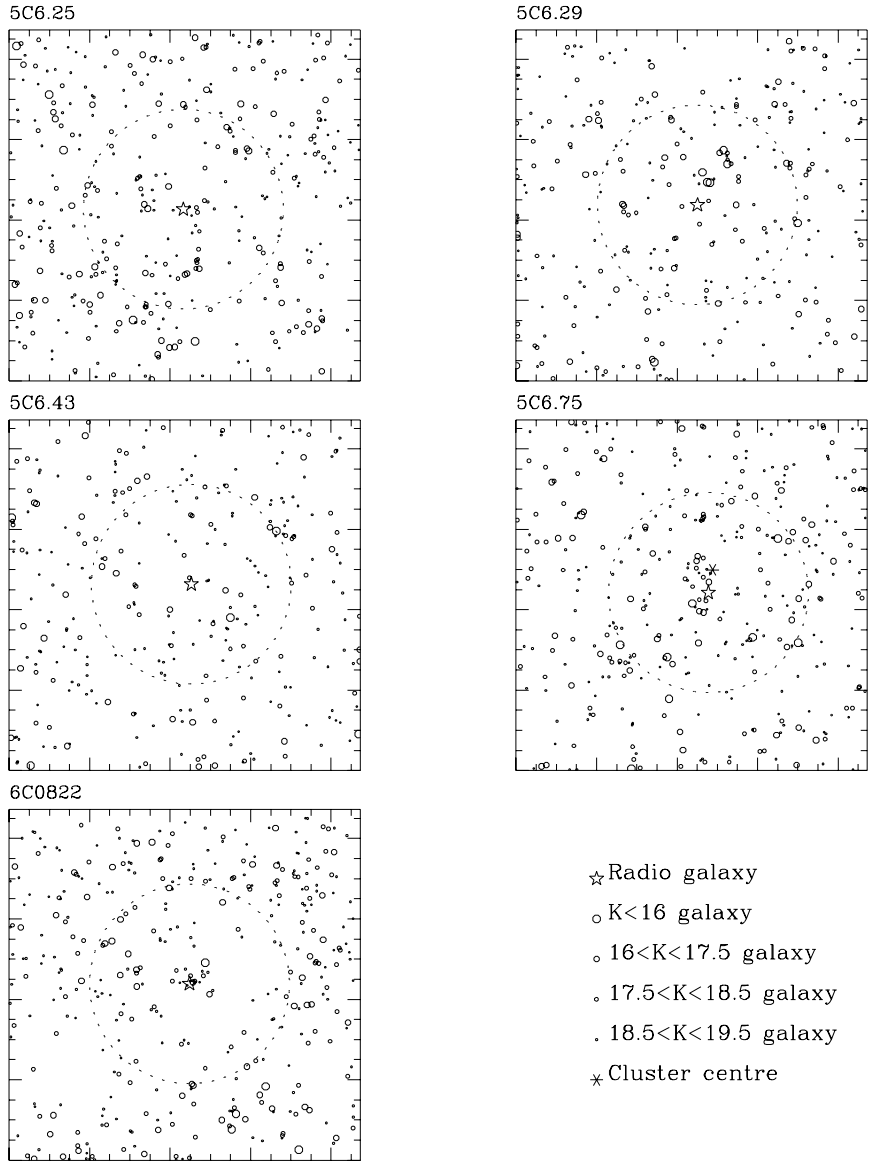,width=170mm}
\caption{Positions of all $K^{\prime}\leq 19.5$
galaxies on the five Omega fields, each $5.75\times 5.75$ arcmin. The short-dashed circles show a 1.5 arcmin radius around each radio galaxy. The best-fit
cluster centre on the 5C6.75 field is shown by an asterisk.}
\end{figure}
\twocolumn

 Following Kepner et al. (1998) we assume $R_c=0.1$ $h^{-1}$ Mpc (where $h$ is Hubble's constant in units of 100 km $\rm s^{-1}Mpc^{-1}$) and $\alpha=0.75$, parameters in the
middle of the observed ranges for both local and distant clusters
 (Girardi et al. 1995; Lubin and Postman 1996), and $r_{max}=1.0 h^{-1}$ Mpc.
With $q_0=0.05$, a proper distance $1.0h^{-1}$ Mpc at $0.7\leq z\leq 0.8$ corresponds to  3.38--3.57 arcmin, so the Omega field of view
would contain almost the entirety of a cluster centred on a radio galaxy. 

Our model for field galaxies incorporates luminosity functions with a much steeper faint-end slope for 
blue than for red galaxies (see Roche and Eales 1998), but this may not be appropriate for rich clusters, which contain a much higher proportion of early-types and many low
luminosity red galaxies. For example, Oemler et al. (1997) estimated that only 12 per cent of the galaxies in two high density Abell class 1 clusters at $z\sim 0.4$ were much bluer than passively evolving ellipticals.
To take this into account, when modelling the cluster we assumed
the k+e-corrections of an evolving S0 model for a large proportion
($78$ per cent) of the galaxies in the steeper luminosity functions
corresponding to Sab and later types,
to give a total red galaxy fraction $\sim 88$ per cent.

Kepner et al. (1998) parameterized the richness of model clusters  in terms of 
the total luminosity ($L_{cl}$,  in units of $L^*$) within the assumed $r_{max}$ of $1 h^{-1}$ Mpc, and  estimated that this related to the Abell richness as
$N_A\simeq {2\over 3} L_{cl}$. We determine $L_{cl}$ for our model clusters
by summing the luminosity of the cluster galaxies (to a faint limit  $K^{\prime}=22$) relative to $L^*$ (hence
$L_{cl}$ is not affected by luminosity evolution).

For modelled clusters at the redshifts of the radio galaxies, 
we estimate that $\sim 0.7$--$0.8L_{cl}$ cluster members will be visible to the completeness limit
of $K^{\prime}=19.5$, about half
of these within 1.5 arcmin of the cluster centre.
 At $z\simeq 0.75$, an $L^*$ elliptical ($M_{K^{\prime}}=-24.85$
for h=0.5) evolving as in our PLE model would
have an apparent magnitude $K^{\prime}\simeq18.0$ and a typical brightest cluster galaxy
($M_{K^{\prime}}=-26.49$ for $h=0.5$, Collins and Mann 1998) would have
 $K^{\prime}\simeq 16.4$, so any galaxies 
associated with the radio galaxies will almost certainly be fainter than $K^{\prime}=16$. 
 We first estimate the cluster environment of the radio galaxies simply
by counting the number of $16\leq K^{\prime}\leq 19.5$ galaxies within 1.5
arcmin of the radiogalaxy positions and comparing with modelled clusters.

\subsection{Excess galaxies within 1.5 arcmin}

Table 2 gives the total number of $16\leq K^{\prime}\leq 19.5$ galaxies within 1.5 arcmin of each of the radio galaxy positions (including the radio galaxies
themselves),  $N(<1.5)$. The background galaxy density for each field is
estimated by dividing the number of galaxies $>1.5$ arcmin from the radio galaxy, $N(>1.5)$ by the area of the part of the field
 $>1.5$ arcmin from the radio galaxy, $A(>1.5)$.
The excess $E$ above the background of galaxies within 1.5 arcmin of the radio galaxy is first estimated by subtracting 
\begin{equation}
E(<1.5)=(N(<1.5)-{\pi(1.5)^2\over A(>1.5)}N(>1.5)
\end{equation}
 However, if the radio galaxy does lie in a rich cluster, some cluster members -- about half if the profile follows equation (1) -- will lie
$>1.5$ arcmin from the cluster centre, causing the background density to be overestimated. For
a modelled cluster of richness $L_{cl}$, we estimate a corrected background
density by subtracting the number of model cluster galaxies $>1.5$ arcmin from the cluster centre,
$C(>1.5)$, from the observed $N(>1.5$). This leads to a higher, corrected estimate
\begin{equation}
E(<1.5)_{corr}= N(<1.5)-{\pi(1.5)^2\over A(>1.5)}[N(>1.5)-C(>1.5)] 
\end{equation}
 The cluster richness is then estimated
as the value of $L_{cl}$ for which 
the number of galaxies within 1.5 arcmin of the centre of a modelled 
$L_{cl}$ cluster, $C(<1.5)$, is equal to $E(<1.5)_{corr}$ with the background correction from the same model cluster. For these fields, the background correction effectively increases
the estimates of $L_{cl}$ by $\sim 35$ per cent. 6C0822
actually gave a negative $E(<1.5)$ -- it may coincide with a  void in the distribution of nearer galaxies, so to reduce the effects of this we compute the excess for a smaller radius of 1.0 arcmin.

\begin{table}
\caption{The number of $16\leq K^{\prime} \leq 19.5$ galaxies within 1.5 arcmin
of the radio galaxy positions (within 1.0 arcmin in the case of 6C0822), the 
excess above the background density (equation 2, with errors $\surd N_{gal}$), the excess 
corrected for the presence of a cluster (equation 3), and the estimated total luminosity $L_{cl}$
of a cluster centred on the radio galaxy.}
 \begin{tabular}{lcccc}
\hline
Galaxy    & $N_{gal}$ & $E(<1.5)$ & $E(<1.5)_{corr}$ & $L_{cl}$ \\
\smallskip
5C6.25  &  82 &  $11.5\pm9.1$ & $15.8\pm 12.5$ &  $40\pm 32$\\           
5C6.29  &  69 &  $8.2\pm8.3$  & $11.5\pm 11.6$ &  $30\pm 31$\\
5C6.43  &  62 &  $4.1\pm7.9$  & $5.4\pm 10.4$  & $15\pm 29$ \\
5C6.75  &  98 &  $33.5\pm 9.9$ & $45.0\pm 13.3$ & $125\pm 37$\\
6C0822  &  36 & $1.6\pm 6.0$ & $2.0\pm 7.5$ &  $8\pm 28$ \\
\hline
\end{tabular}
\end{table}

Of the 5 radio galaxies, only 5C6.75 shows a significant 
excess of nearby galaxies over the scales expected for a rich cluster. This excess 
would correspond to a cluster of Abell richness $N_A\simeq 83\pm 25$, i.e. Abell class 1 or 2. On the basis of $\surd N_{gal}$ statistics the cluster detection is $3.34\sigma$, although this may be an overestimate due to the clustering of background galaxies -- see Sections 4.2 and 5.3 for estimates of the 
significance which take this into account.

Of the other galaxies, 5C6.25 
appeared to the eye to be in a large-scale association, but the estimated richness is probably too low for this to be described as even an Abell 0 cluster,
and 6C0822 appear to be in an association of a few 
galaxies, but the lack of any excess of galaxies over cluster scales of $\sim 1$ arcmin suggests this is a small isolated group rather than a true cluster.
For the five radio galaxies, the mean $L_{cl}$ is $43.6\pm21.1$, corresponding 
to a mean Abell richness $29\pm 14$ -- this is
discussed in Section 7.1.

\section{Cluster Detection}

\subsection{Method}

We now apply a second method of searching for clusters, which makes use of the full model cluster profile of equation (1). This has the advantages of (i) providing an estimate of the cluster centroid position -- assuming clusters are centred exactly on the radio galaxies is likely to underestimate the richness if there is an offset, (ii) providing an estimate of the richness which takes into account the full radial distribution and should therefore 
give greater 
statistical power than a simple summation to a fixed radius, (iii) deriving errors from the real distribution of galaxies, rather than by assuming $\surd N$ statistics, and thus verifying that any cluster detections remain significant    after taking into account the clustering of background galaxies.

We search the regions near the radio galaxies for clusters using a 
cluster detection routine, which is essentially a simplified version
of the Adaptive Matched Filter technique described by Kepner et al. (1998).
The profile of a cluster of chosen redshift $z$ and richness $L_{cl}$ is modelled as in
Section 3.2. For a chosen position ($x$,$y$), the number
of $16\leq K^{\prime}\leq 19.5$ galaxies are counted in annular bins of
$\Delta(\theta)=10$ arcsec centred on ($x$,$y$). The number of galaxies in a 
 model cluster, placed at ($x$,$y$), are counted in the same
annular bins, with areas missed off the edges of the field or lost due
to `holes' taken into account.
The model and observed counts are then compared by summing
\begin{equation}
\chi^2(x,y,L_{cl},z)={1\over n_b-1}\sum_{i=1}^{i=n_b}
{(C_i-N_i+ A_i\rho_{back})^2\over A_i\rho}
\end{equation}
where $C_i$ is the number of galaxies in bin $i$ in the modelled cluster,
$N_i$ is the observed number of galaxies in bin $i$, $A_i$ is the area 
of bin $i$ (taking into account any area lost due to holes or the field edges), $\rho$ the mean surface density of galaxies on the whole field and $\rho_{back}$ a corrected background density  
(the observed number of galaxies on the field 
minus the total number that would be in the modelled cluster, divided by the 
field area).

The $\chi^2$ statistic is computed for a grid of ($x$,$y$) positions on each field, for a model cluster with 
$z$ equal to the radio galaxy redshift and $L_{cl}=125$, the divide between Abell classes 1 and 2 and the approximate richness of the 5C6.75 cluster.
For each field $\sigma(\chi^2)$ is computed as the scatter in $\chi^2$ values
between 4900 ($x$,$y$) positions, thus 
 taking into account both the non-independence of the bins and the greater than
Poissonian variations in the galaxy surface density due to clustering.
Candidate cluster positions are identified as minima in the $\chi^2$ map, where
$\chi^2$ is significantly, in terms of $\sigma(\chi^2)$, lower than the mean for the whole field.

\subsection{Results}

The strongest detection of a cluster was centred at pixel co-ordinate (488,499) on the
5C6.75 field, where the $\chi^2$ for a  $L_{cl}=125$ cluster profile was 
$3.06\sigma$ below the mean for all positions on the field. This confirms that
the cluster remains a $3\sigma$ detection after taking into account the  increased variations in the background density from the clustering of other galaxies on the field. There are no $3\sigma$ detections of clusters within 1.5 arcmin of any of the other four radio galaxies.

The best-fit cluster position, shown on Figure 4, is offset by 23 arcsec ($\simeq 0.11 h^{-1}$ Mpc) from the radio galaxy (478,442). However, running 
the cluster detection routine on 10
simulated $L_{cl}=128$ clusters, generated by distributing galaxies within a probability distribution following the profile from Equation 1 on top of a randomly distributed background gives the 
RMS error in the best-fit position as $\sim 21$ arcsec, so this offset is
not significant. 

There are 99 galaxies with $16\leq K^{\prime}\leq 19.5$ within 1.5 arcmin of the fitted cluster centre, an excess of 
$34.4\pm 9.9$ above the background or $46.6\pm 13.4$ above the
corrected background, giving $L_{cl}=128\pm 37$ (hence
Abell richness $N_A=85\pm 25$), almost identical to the estimate when the radio galaxy is assumed to be the centre.
The richness can also be estimated using the cluster detection routine, placing a model cluster at the best-fit position and finding the model $L_{cl}$ which
mimimizes $\chi^2$. This gives 
$L_{cl}= 136\pm 26$ (error from the simulation), consistent with the first estimate. 

Figure 5 compares the observed cluster profile with the 
$L_{cl}=128$, $R_{c}=0.1$ $h^{-1}$ Mpc model.
 Allowing
both $L_{cl}$ and $R_{c}$ to vary gives a best-fitting core radius as 
$R_c=0.12\pm 0.03 h^{-1}$ Mpc. The good agreement of the observed cluster profile and its best-fit $R_c$ with the model profile confirms that the distribution of galaxies near 5C6.75 is at least consistent with an
Abell 1/2 cluster, similar in profile to those in the local Universe, at the radio galaxy redshift.

\begin{figure}
\psfig{file=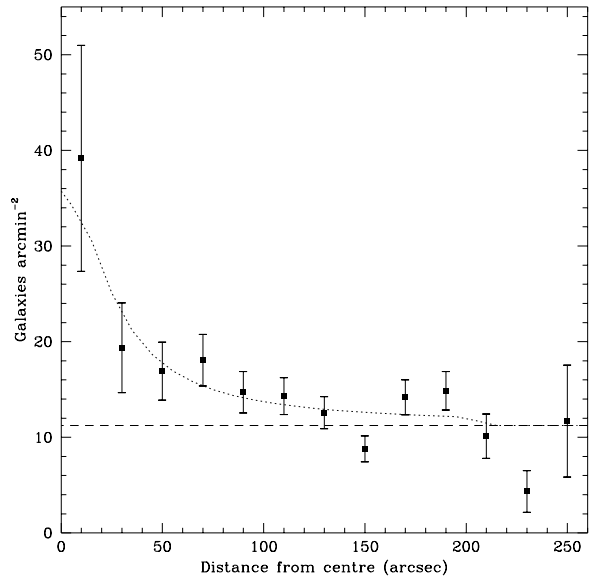,width=85mm}
\caption{Radial density profile of the 5C6.75 cluster, with $\surd N$ error
 bars, compared to (dotted line) a model cluster profile (equation 1) with $L_{cl}=128$
and $R_{c}=0.1$ $h^{-1}$ Mpc.}
\end{figure}

\section{The angular correlation function}

\subsection{Calculating $\omega(\theta)$}

We investigate the clustering of all the galaxies on these five fields
by calculating the angular correlation function, $\omega(\theta)$.
Our data may not give an entirely unbiased
estimate of the field galaxy $\omega(\theta)$, due to the
presence of known radio galaxies with strong clustering around at least one.
However, the great majority of detected galaxies will not be associated with the
radio sources, so the resulting bias may be very small. For now, we analyse these images as normal field samples, with no distinction between radio
and other galaxies, but investigate the effect of the 5C6.75 cluster at the end of Section 5.2.

For each field, $\omega(\theta)$ is calculated for all detections classed as galaxies to a series of faint magnitude limits. On a field with  $N_{g}$ 
galaxies brighter than the chosen limit, there will be
${1\over2}N_{g}(N_{g}-1)$ possible galaxy-galaxy pairs, which 
counted in bins of separation of width $\Delta ({\rm log}~\theta)=0.2$,
giving a function $N_{gg}(\theta_i)$.
For each field $N_r=10000$ points are placed at random over the area
covered by real data, i.e. avoiding any holes, and
 the separations of the $N_{g}N_{r}$ galaxy-random pairs, taking the real galaxies as the centres, are similarly counted, giving $N_{gr}(\theta_i)$. In addition, the separations of the ${1\over2}N_r(N_r-1)$ random-random
pairs are counted in bins to give $N_{rr}(\theta_i)$. 

Defining $DD=N_{gg}(\theta_i)$, and $DR$ and $RR$ as the galaxy-random
and random-random counts normalized to have the same summation over all 
$\theta$ as $DR$, i.e.
$DR={(N_g-1)\over 2N_r} N_{gr}(\theta_i)$ and
$RR={N_g(N_g-1)\over N_R(N_r-1)} N_{rr}(\theta_i)$,
we calculate $\omega(\theta)$ for each $\theta_i$ bin using the  Landy and Szalay (1993) estimator,
\begin{equation}
\omega(\theta_i)={DD-2DR+RR\over RR}
\end{equation}
The $\omega(\theta_i)$ of the five fields are averaged at each $\theta_i$
point to give a mean $\omega(\theta)$, shown on Figure 6 for
magnitude limits $K^{\prime}=18.5$--20.0, with error bars from the field-to-field 
scatter.

If the real galaxy $\omega(\theta)$ is of the form $A\theta^{-\delta}$, the observed $\omega(\theta)$ will follow the form $\omega(\theta)=
A(\theta^{-\delta} - C)$, with amplitude $A$ (defined here at a one-degree
separation), and a negative offset $AC$ (the integral
constraint) resulting from the restricted area of the
observation, which can be estimated by doubly
integrating an assumed true $\omega(\theta)$ over the field area $\Omega$,   
\begin{equation}
AC={1\over \Omega^2}\int\int \omega(\theta) d\Omega_1 d\Omega_2
\end{equation}
Using the random-random
correlation, this calculation can be done numerically --
\begin{equation}
C={\sum N_{rr}(\theta) \theta^{-\delta}\over \Sigma N_{rr}(\theta)}
\end{equation}
Assuming $\delta=0.8$, in
agreement with most observations including those in the $K$-band at brighter limits (Baugh et al. 1996), $C=14.84$ for these fields. The $\omega(\theta)$ amplitude $A$ is then estimated by fitting
$A(\theta^{-0.8}-14.84)$ to the mean $\omega(\theta)$ at separations
$2<\theta<200$ arcsec.
The error on $A$ is estimated by fitting the 
same function to the $\omega(\theta)$ of the five individual fields and 
determining the scatter between the individual field amplitudes.

\subsection{$\omega(\theta)$ results}

Figure 6 shows the observed $\omega(\theta)$ with best-fitting functions of the form   $\omega(\theta)=A(\theta^{-0.8} -14.84)$, with the amplitudes $A$ in Table 3. At magnitude limits $K^{\prime}=18.5$--20.0, there is a  
significant ($\sim 4\sigma)$ detection of galaxy clustering, 
with the individual $\Delta({\rm log}~\theta)=0.2$
bins showing a significant ($\sim 3\sigma$) and positive signal out to separations 
 ${\rm log}~\theta=-2.25=20$ arcsec. The result for $K^{\prime}=20$ may be less reliable due to incompleteness at $K^{\prime}>19.5$ (Section 2.3), whereas at limits of  $K^{\prime}=18.0$ and brighter, there are an insufficient number of galaxies for a significant detection of clustering.

 The first two plotted points, corresponding to $1.26\leq \theta\leq 3.17$
arcsec, lie above the fitted power-laws, indicating an excess of close pairs of galaxies relative to the clustering seen at larger scales. 
 In Paper I, the large number of 2--3 arcsec pairs on the Redeye camera fields was attributed to the effects of galaxy mergers, but it was also suggested  that
$\omega(\theta)$ from $K$-band surveys might be steeper than $\theta^{-0.8}$, 
due to a particularly steep two-point correlation function, $\xi(r)$, for
giant ellipticals.
Loveday et al. (1995) measured a $\xi(r)$ slope
$\gamma=1.87\pm 0.07$ for E/S0 galaxies and Guzzo et al. (1997) a more
extreme $\gamma=2.05\pm 0.09$ for higher luminosity E/S0s.
The latter estimate corresponds to $\omega(\theta)\propto \theta^{-1.05}$, which we assume as an extreme upper limit
to the possible range of slopes, and fit the observed $\omega(\theta)$ with `$A_m(\theta^{-1.05}-C_m)$', where  $C_m=38.19$ for a $\theta^{-1.05}$ power-law (from equation 7).

Figure 6 shows this steeper power-law with the best-fitting amplitudes
($A_m=3.93\pm0.39\times 10^{-4}$, $2.88\pm0.62 \times 10^{-4}$,
$3.132\pm 0.49\times 10^{-4}$, $2.49\pm 0.46\times 10^{-4}$ at limits $K^{\prime}=18.5,19.0,19.5$ and 20.0 respectively). Changing from
$\theta^{-0.8}$ to $\theta^{-1.05}$ does not get even half-way to
fitting the small-scale excess (except perhaps at the $K^{prime}=20$ limit where close-pair detection is likely to suffer incompleteness), suggesting that the
dominant contribution to this is from 
interacting galaxies rather than  steeper E/S0 clustering.
The excess of close pairs is investigated in Section 6 and discussed in Section 7.3.

 A fraction of (randomly distributed) stars $f_s$ within the galaxy sample will reduce the observed $\omega(\theta)$ at all angles by a factor of
$(1-f_s)^{-2}$.  For each magnitude limit, we estimate, using the star-count model of Section 2.3,  the fraction $f_s$ of faint ($K>17$) 
stars contaminating the sample of $N_{gal}$ objects. Table 2 gives $f_{s}$ 
and $\omega(\theta)$ amplitudes
$A_{corr}$ corrected for star contamination by multiplying the fitted $A$ by
$(1-f_s)^{-2}$, with errors combining in quadrature the $\omega(\theta)$
errors with the correction error (derived from the 21 per cent 
uncertainty in the star-count normalization).

\begin{table}
\caption{Observed $\omega(\theta)$ amplitudes ($A$), in units of $10^{-4}$ at $1^{\circ}$, the number of galaxies ($N_{gal}$), the estimated fraction of contaminating stars $f_s$ and the $\omega(\theta)$ amplitudes corrected for star-contamination $A_{corr}$, 
 at a series of $K^{\prime}$ magnitude faint limits.}
\begin{tabular}{lcccc}
\hline
$K^{\prime}$ limit & $A$ & $N_{gal}$ & $f_s$ & $A_{corr}$ \\
18.5 & $12.77\pm 1.84$ & 807 & $0.180\pm 0.038$ & $21.94\pm 3.25$ \\
19.0 & $11.28\pm 2.98$ & 1117 & $0.164\pm 0.034$ & $16.14\pm 4.47$ \\
19.5 & $13.25\pm 2.27$ & 1626 & $0.150\pm 0.032$ & $18.34\pm 3.43$ \\
20.0 & $11.20\pm 2.19$ & 2110 & $0.148\pm 0.031$ & $15.43\pm 3.22$ \\
\hline
\end{tabular}
\end{table}

It might be thought that the cluster on the 5C6.75 field
 would upwardly bias our estimate of the field galaxy $\omega(\theta)$, but repeating the analysis with galaxies within  1.5 arcmin of the cluster centre  excluded produces little change in the results (in units of $10^{-4}$, the
uncorrected $\omega(\theta)$ amplitudes become $10.10\pm2.55$, $10.09\pm 2.53$,
 $15.13\pm 2.85$ and $11.21\pm 2.73$ at $K^{\prime}=18.5,19.0,19.5$ and 20.0
respectively). Although the $\xi(r)$ amplitude in the core of Abell 1/2 clusters
may be several times higher than for
field galaxies (see e.g. Hill and Lilly 1991), the excess  galaxies within 1.5 arcmin of 5C6.75
amount to only 13 per cent of all galaxies on the CCD frame, which in turn is
only one-fifth of our dataset, so the effect
of their clustering on $\omega(\theta)$ is greatly diluted.
\onecolumn
\begin{figure}
\psfig{file=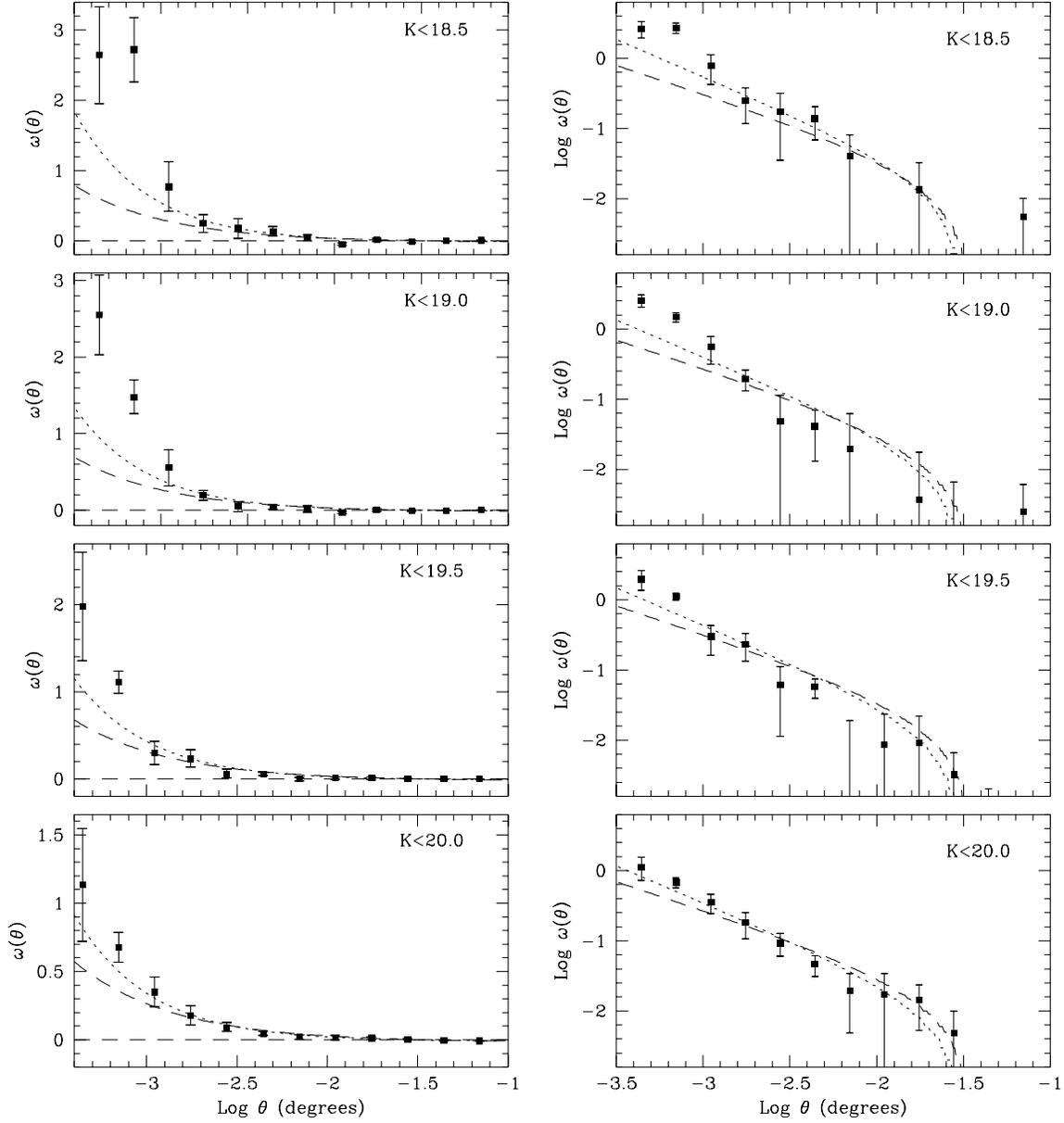,width=170mm}
\caption{Observed $\omega(\theta)$ for galaxies brighter than 
$K^{\prime}=18.5$, 19.0, 19.5 and 20.0, as log-linear (left) and log-log (right)
plots. The dashed lines show the best-fit $\omega(\theta)=A(\theta^{-0.8} -1.84$
functions and the dotted lines the best-fit functions for a 
maximum slope, of the form $A_m(\theta^{-1.05}-C)$.}
\end{figure}

\twocolumn

Figure 7 shows the corrected and uncorrected $\omega(\theta)$ amplitudes as a function of
$K^{\prime}$ limit, from this survey and others in the $K$ band, and models described in Section 5.4. Note that the Paper I and Carlsberg et al. (1997) results are plotted without a correction for star-contamination (Baugh et al.
1996 are at sufficiently bright magnitudes that star-galaxy separation
should be reliable). At  $K=19.5$--20.0, our uncorrected amplitudes are consistent with Paper I. 

\begin{figure}
\psfig{file=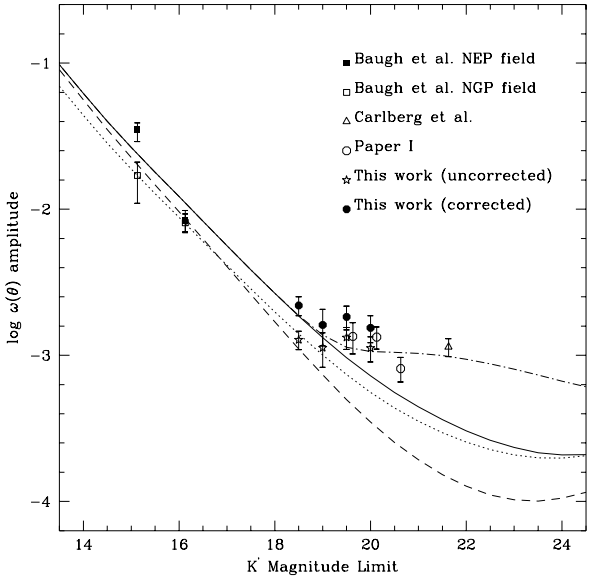,width=85mm}
\caption{The $\omega(\theta)$ amplitude for galaxies on our five fields,
with and without corrections for star-contamination, against $K^{\prime}$ magnitude limit, compared with $\omega(\theta)$ amplitudes from the $K$-band surveys of Baugh et al. (1996), Carlberg et al. (1997) and Paper I.
The solid line shows a PLE model (Roche and Eales 1998) with $r_0=8.4
h^{-1}$ Mpc for early-type galaxies and stable clustering ($\epsilon=0$),
the dashed line the same PLE model with $\epsilon=1.2$ evolution of the
clustering, the dot-dash line the same model with increased clustering at $z>1.5$ (see Section 7.2), and the dotted line the $\epsilon=0$ PLE model with the $r_0$ from Roche et al. (1996),  
$r_0=5.9 h^{-1}$ Mpc for E/S0 galaxies.}
\end{figure}
\subsection{Effect on detection of rich clusters}
Having directly measured the galaxy clustering on these fields we can  estimate its effect on the detection of rich clusters around the radio galaxies. Considering cells of area $\Omega$, with a mean background density of $\langle N \rangle$ galaxies in each, clustering will increase the variance $\mu_2$ of the counts in these cells
 as
\begin{equation}
\mu_2=\langle N \rangle + =\langle N \rangle^2 {1\over \Omega^2} \int\int \omega(\theta) d\Omega_1 d\Omega_2
\end{equation}
which for $\omega(\theta)=A\theta^{-0.8}$ and a circular area of radius 1.5 arcmin gives $\mu_2=25.6 A\langle N \rangle^2$. The background density for 
the 5C6.75 field was estimated as $\langle N \rangle=53$ and the uncorrected (as the background density will include contaminating stars) 
$A=1.325\times 10^{-3}$, giving a background $\sigma=\surd(53+95.3)=12.2$.
The significance of the excess of galaxies within 1.5 arcmin of 5C6.75 is then reduced from $3.34\sigma$ to ${33.5\over 12.2}=2.75\sigma$. This is slightly lower than the $3.04\sigma$ significance from the cluster detection routine, which is presumably more statistically powerful as it takes into account the full cluster profile.

 We conclude that the effects of the clustering of unassociated  galaxies on our investigation of radio galaxy cluster environments
are non-negligible but relatively small, a $\sim 20$ per cent increase in statistical errors, and that the detection of the
5C6.75 cluster remains close to $3\sigma$.

\subsection {Comparison with models}

The modelling of $\omega(\theta)$ amplitudes is disussed in detail elsewhere
(e.g. Roche et al. 1996; Roche and Eales 1998) and will be described only briefly here. Essentially, the 3D two-point correlation function
for the galaxies is parameterized as
\begin{equation}   
\xi(r,z)=(r/r_0)^{-\gamma}(1+z)^{-(3+\epsilon)}
\end{equation}
 where $r_0$ normalizes the strength of clustering
at $z=0$, $\gamma$ is the slope and $\epsilon$ the clustering 
evolution relative to the $\epsilon=0$ stable clustering model. This is 
integrated over a galaxy redshift distribution, $N(z)$, using 
Limber's formula (e.g. Phillipps et al. 1978).
Here $N(z)$ is given by the PLE model of Section 2.3.

Roche et al. (1996) assumed $\gamma=1.8$ for all galaxy types, with
$r_0=5.9$ $h^{-1}$ Mpc for E/S0 galaxies
and $r_0=4.4$ $h^{-1}$ Mpc for later types, with an additional luminosity
weighting so that galaxies with $z=0$ blue-band absolute magnitudes $M_B>-20.5$
(for $H_0=50$ km $\rm s^{-1}Mpc^{-1}$)  are half as clustered as those of higher luminosity (see Loveday et al. 1995).
With stable clustering and $L^*$ evolution, this model predicted a $\omega(\theta)$ scaling  consistent with observations in the blue-band, but  appeared to underpredict the 
$\omega(\theta)$ amplitudes from $K$-band surveys at $K\simeq 19.5$--21.5 (Paper I). The
 results from the deep $K$-band surveys, in which early-type galaxies will be much more prominent, were better fitted by a model with even stronger clustering for early-type galaxies.

The $\omega(\theta)$ scaling from a large $R$-band 
survey (Roche and Eales 1998) appeared
well-fitted by a
PLE model with $\gamma=1.8$, $r_0=8.4$ $h^{-1}$ Mpc for E/S0 galaxies and
$r_0=4.2$ $h^{-1}$ Mpc for spirals and irregulars, again with the same luminosity weighting, and $\epsilon\simeq 0$, whereas
clustering evolution of $\epsilon\sim 1.2$   
underpredicted the $\omega(\theta)$ amplitudes. 
On Figure 7, the same PLE model with $\epsilon=0$ but the $r_0$ from Roche et al. (1996) underpredicts the
$\omega(\theta)$ amplitudes from this survey and is rejected 
 after star-contamination is taken into account. With the $r_0$ from Roche and Eales (1998), the PLE model with $\epsilon=0$ gives higher $\omega(\theta)$ amplitudes more consistent with the observations. The same model with  clustering evolution of $\epsilon=1.2$  greatly underpredicts the $\omega(\theta)$ of deep $K$-band surveys, and at $K^{\prime}=19.5$ is rejected by $>3\sigma$ even before the
correction for star-contamination. 

At our fainter limits of $K^{\prime}= 19.5$--20.0, there may be some evidence that the $\omega(\theta)$ amplitude falls less steeply than the  $\epsilon= 0$ model,  and 
the amplitude from the small area studied by Carlberg et al. (1997) $K^{\prime}=21.5$ suggests that the $\omega(\theta)$ scaling actually levels out.  
This might indicate an increase in the  clustering of red galaxies at the highest redshifts ($z\geq 1.5$), perhaps 
related to the enhanced clustering reported for Hubble Deep Field  galaxies at
$z\geq 2.4$ (Magliocchetti and Maddox 1998), but more data is needed to confirm this. If we assume the levelling out of the $\omega(\theta)$ scaling is genuine,  it can be fitted by the
 $\epsilon=0$ PLE model with an increase in clustering
to $r_0=16.15 h^{-1}$ Mpc (see Section 7.2) for all $z>1.5$ galaxies, as shown on Figure 6.

In summary, the $\omega(\theta)$ amplitudes from this survey are consistent with previous $K$-band observations, favouring both strong
($r_0\sim 8.4$ $h^{-1}$ Mpc) clustering for E/S0 galaxies and clustering 
remaining stable ($\epsilon\sim 0$) to $z\sim
1.5$; if anything the clustering is stronger than an $\epsilon=0$ model at the highest redshifts.
We discuss this further in Section 7.2. 
  
\section{Close pairs of galaxies}

\subsection{Method}

The $\omega(\theta)$ from this data exceeded the fitted power-law at small separations of $1.26\leq  \theta
\leq  3.17$ (Figure 6), indicating that there is an excess of close pairs of galaxies. As in paper I, we quantify this using a
method described by Woods, Fahlman and Richer (1995). A probability $P$ of occurring by chance (in a random distribution
of galaxies) is estimated for each  galaxy-galaxy pair, as 
\begin{equation}
$$P=\int^{\theta}_{\beta}{\rm exp}(-\pi\rho\alpha^2) d\alpha$$
\end{equation}
where $\rho$ is the surface density of galaxies brighter in apparent magnitude
than the fainter galaxy of the pair, $\theta$ is the pair separation and
$\beta$ an angular separation cut-off below which individual objects cannot
be resolved.

We take $\beta=1$ arcsec and  find the area around each galaxy 
by counting randomly distributed points in annular bins rather than assuming $\pi r^2$, thus taking
into account field edges and holes. For all 1629 objects classed as galaxies 
to the estimated completeness limit $K^{\prime}=19.5$, 
the pairs with $P\leq 0.05$ are counted in $\Delta(\theta)=0.5$ arcsec
bins of separation. We then perform an identical analysis on 25 randomized datasets in 
which the same number of `galaxies', with the same magnitudes, are redistributed randomly over the field areas. The pairs counts from the randomized dataset and 
averaged, and their scatter used to derive error bars for a single dataset.

\begin{figure}
\psfig{file=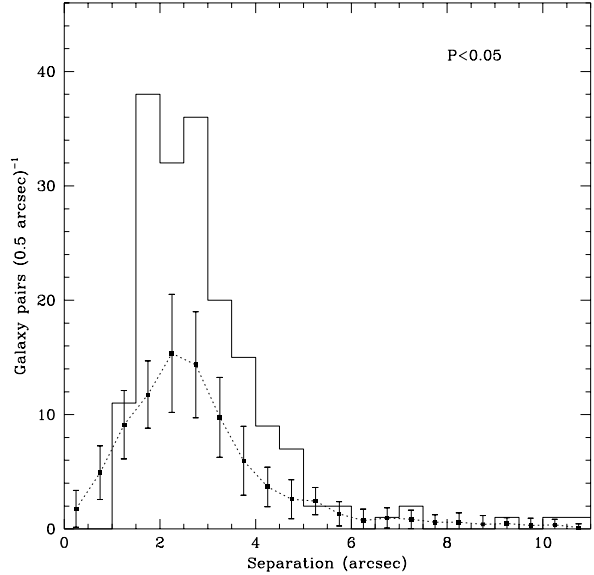,width=85mm}
\caption{ Histograms of the observed numbers of
 galaxy-galaxy pairs (to $K^{\prime}=19.5$) with probability
$P<0.05$. The dotted line shows the pair count from 
simulated random distributions of galaxies, with error bars from
the simulations.
}
\end{figure}

\subsection{Results}
Figure 8 shows a histogram for the $P\leq 0.05$ pair counts compared to the
random expectation. Over the $1.5<\theta<5.0$ arcsec range, 
there are 157 pairs with
$P\leq 0.05$ compared to $63.48\pm 10.57$ expected for a random
distribution of galaxies. The excess above random, assuming the error to scale as $\surd N$, consists of $93.52\pm 16.62$ pairs. 

One one field, every pair was examined by eye, and only 2 out of 38 appeared spurious (caused by irregular outer regions of bright galaxies), while the other 36 appeared to be genuine galaxy pairs, in many cases with signs of interaction. There are 15 $1.5<\theta<5.0$ arcsec pairs within 1.5 arcmin of the cluster radio galaxy 5C6.75. This can be compared to 3.16 expected for a random distribution, increasing to 7.8 or 10.8 if, respectively, the pairs/randoms and pairs/galaxies ratios for this area are the same as for the whole dataset.
 Hence our discovered
 cluster may contribute a few of the excess pairs, but increases the estimate of the fraction of galaxies in pairs by only $\sim {4.2\over (157-4.2)}=2.7\pm 1.3$ per cent. 

No $\theta<1$ arcsec pairs are counted, as with the $\rm FWHM\simeq 1.13$ arcsec resolution they would be merged into single detections.
The observed number of $1.0<\theta<1.5$ arcsec pairs is approximately the random expectation, suggesting (by comparison with the number of $1.5<\theta<5.0$ arcsec pairs) that only $\sim 40$ per cent of the true number are resolved. The number of close pairs missed due to image merging is 
estimated  by assuming that all $\theta>1.5$ arcsec pairs are resolved and that
the small-scale $\omega(\theta)$ is still a $\theta^{-0.8}$ power-law, which gives the true number of pairs above the random
expectation at $\theta<1.5$ arcsec as 0.3068 the number at $1.5<\theta<5.0$ arcsec.

To estimate the number of physically merging or interacting galaxies, we must subtract from the observed number of pairs, the number expected from normal galaxy clustering, which will be higher than the random expectation. Clustering with a $\theta^{-0.8}$
power-law of amplitude $\alpha$ at 1 arcsec increases the ratio
of $\theta<5$ arcsec pairs to the random expectation by  
\begin{equation}
{\int_0^{5}2\pi \theta\alpha\theta^{-0.8} d\theta\over \pi(5)^2}
={2\pi\alpha\over 25\pi}{\theta^{1.2}\over 1.2}=0.46\alpha
\end{equation}
 Our measured $\omega(\theta)$ amplitude at the $K^{\prime}=19.5$ limit corresponds to
$\alpha=0.93\pm 0.16$ without correction for
star-contamination or $\alpha=1.28\pm 0.24$ with correction, and 
the random expectation for $\theta<5$ arcsec pairs is 79.28.
Hence,  if we first neglect the star-contamination, the excess of $\theta<5$ arcsec  pairs
above the expectation from the larger scale $\omega(\theta)$ 
amounts to
$1.3086\times (93.52\pm 16.62)-0.46\times (0.93\pm 0.16)\times 
79.28=88.46\pm 22.52$. 
As each pair consists
of two galaxies, the fraction of the 1629 galaxies in merging/interacting
pairs is then $(2\times 88.46\pm 22.52)/1629=10.9\pm 2.8$ per cent.

Taking into account the star contamination, the excess of
$\theta<5$ arcsec pairs is reduced to
$1.3086\times (93.52\pm 16.62)-0.46\times (1.28\pm 0.23)\times 
79.28=75.70\pm 23.31$, but the number of real galaxies is also reduced, to 1382,
leading to a very similar estimate of the merging/interacting fraction as
 $(2\times 75.70\pm 23.31)/1382=11.0\pm 3.4$ per cent.
This estimate  is discussed further in Section 7.3.

\section{Discussion}

\subsection{The radio galaxy clustering environment}

At low redshifts, the most radioluminous (FRII) radio galaxies  tend not to be found in galaxy clusters, whereas the much lower luminosity FRI sources often are. This scenario changes at relatively moderate redshifts -- Hill and Lilly (1991) found that 43 
radio galaxies at $0.35<z<0.55$, with a very wide  range of radio luminosities, were distributed in
similar numbers over field, Abell 0 and Abell 1 cluster environments. The mean environment of the 43, quantified in terms  of $B_{gg}$, the normalization of $\xi(r)$ at $r=0.5 h^{-1}$ Mpc, was estimated as  $B_{gg}=291\pm 45$, 
approximately an Abell 0 cluster, with
no strong dependence on radio luminosity -- for the most luminous (3C)
subsample the mean environment was $B_{gg}=342\pm 96$. It was concluded that the
 mean $B_{gg}$ of luminous (FRII)
radio galaxies (including those of with the radio luminosities of our 7C sample)
increased by a factor $\sim 2.5$ from $z\sim 0$ to $z\sim 0.5$.
We aim to determine whether there is a further increase at $z>0.5$.

Previously,
Bower and Smail (1997) detected by gravitational lensing a cluster of apprximately Abell 2 richness
around a radio-loud QSO (3C336) at $z=0.927$, but found no
significant lensing signal for 
7 other 3C radio galaxies at $0.87<z<1.05$, indicating a mean environment no richer than an Abell 1 cluster. 
 In Paper I, we were able only to set a similar upper limit ($B_{gg}\leq 756$ at $2\sigma$) on the strength of clustering around radio galaxies at $z_{mean}\simeq 1.1$, but here, 
due to the more moderate ($0.7<z<0.8$) radio galaxy redshifts and the larger field size, we are more successful. We detect at $\sim 3\sigma$ significance a cluster of estimated Abell richness $N_A=85\pm 25$ (class 1 or 2), approximately
centred on the
radio galaxy 5C6.75, with a profile consistent with that of a typical present-day rich cluster ($R_c\simeq 0.1$ $h^{-1}$ Mpc, $\alpha\simeq 0.75$ in equation 1) at the radiogalaxy redshift.
Of the other four radio galaxies, two 
appeared to be in groups or structures of lower richness than Abell clusters, and two in field environments.

We estimated the mean environment of our sample of
$0.7<z<0.8$ radio galaxies as Abell richness
$N_A=29\pm 14$. Using large samples of clusters, Lubin and Postman (1997) find
$N_A\simeq $2.5--2.7 $N_0$, where $N_0$ is the core ($<0.25$ $h^{-1}$ Mpc) richness, and Hill and Lilly (1991) fit $B_{gg}\simeq 30 N_0$. Combining these
two relations,  $N_A=29\pm 14$ corresponds to $B_{gg}=335\pm 162$, consistent with the Hill and Lilly (1991) 
estimate for radio galaxies at $0.35<z<0.55$. Hence, 
although a larger  sample will be required  to confirm this, our results together with Paper I and Bower and Smail (1997),
suggest there is little further evolution in the distribution of luminous radiogalaxy environments from $z=0.5$ to $z=1$. 

Wan and Daly (1996) explain the increase in the mean richness of FRII radio galaxy environments between $z=0$ and $z=0.5$ as the result of an increase with time in the typical intra-cluster medium pressure; once the
pressure in a cluster reaches a critical threshold FRII activity is surpressed and only the much less radioluminous FRI outbursts will occur, so FRIIs tend not to be found in clusters at lower redshifts. We infer from this that, firstly, if FRI radio galaxies are not surpressed by intra-cluster pressure, their present-day distribution -- a wide range of field and cluster environments with the mean richness being about that of an Abell class 0 cluster (Prestage and Peacock 1988; Hill and Lilly 1991; Zirbel 1997) -- should reflect the range of environments of massive black holes capable of radio outbursts. Secondly, if at some high redshift, all clusters have an intra-cluster pressure below the threshold for FRII surpression, the distribution of 
FRII environments should then be similar to that of the low redshift FRIs.  

Hence, on the basis of the Wan and Daly (1996) model, the rising trend 
in the mean richness
of FRIIs environments with redshift should approach -- but not exceed -- that of an Abell 0 cluster at high redshifts, and,  as the Abell 0 richness has already been reached at $z\sim 0.5$, there would be little further change 
at $z>0.5$. This is consistent with our observations.

The removal of a strong cluster-environment influence on the activity of radio galaxies beyond $z\sim$0.5--0.6 might also help to explain the  tighter correlation of radio luminosity to host galaxy mass for $z>0.6$ radio galaxies,  compared to  those at lower redshifts
(Eales et al. 1997; Roche, Eales and Rawlings 1998).

\subsection{The galaxy $\omega(\theta)$ amplitude}

At magnitude limits $K^{\prime}=19$--20, we detect galaxy clustering at 
$\sim 4\sigma$ significance. In Paper I we interpreted the relatively high $\omega(\theta)$ amplitude at $K\sim 20$ as evidence that the 
red early-type galaxies were much more clustered than spirals, even at $z\sim 1$, and followed a stable clustering ($\epsilon\sim 0$) model.
This paper's results support these conclusions at better 
statistical significance.
The $K$-band $\omega(\theta)$ scaling is reasonably consistent with a PLE model with
$\epsilon=0$ and $r_0$ of $4.2 h^{-1}$ Mpc for spirals,
$8.4 h^{-1}$ Mpc for E/S0 galaxies. 
 The high $\omega(\theta)$ amplitude and its slow 
decline on going faintward strongly favour $\epsilon\sim 0$ and disfavour clustering evolution of
$\epsilon\sim 1.2$ by $\sim 3\sigma$.

 The larger $R\leq 23.5$ survey of Roche and Eales (1998) found $\omega(\theta)$
to be consistent with the same PLE model as used here, with stable clustering, and Postman et al. (1998) found the $\omega(\theta)$ in
even larger $I<23$ survey best-fitted by an essentially similar model with
$N(z)$ derived from the CFRS, $\epsilon=-0.20\pm 0.17$ and a mean $r_0=5.6\pm 0.23 h^{-1}$ Mpc. 
As most of the higher 
redshift galaxies in a $2.1 \mu \rm m$ survey will be red E/S0s (Figure 3),
whereas the $R$ and $I$-band surveys
will be dominated by later-type galaxies,
the fact that surveys of similar depth in $R$, $I$ and $K^{\prime}$ all favour stable clustering 
 suggests $\epsilon\sim 0$ for the  separate E/S0 and spiral 
populations. The former would then be $\sim 0.5$ dex more clustered at all redshift (to at least 
$z\sim 1.5$),  as suggested by the  Neuschaefer et al. (1997) $\omega(\theta)$ analysis where bulge-profile galaxies on deep HST images remain $\sim 0.5$ dex more clustered 
than disk galaxies from $I\sim 18$ to $I\sim 24$.

 CDM models with $\Omega=1$  
predict $\epsilon\simeq 1$--1.5 for the mass distribution over the redshift range of our survey (Col/'in, Carlberg and Couchmann 1997; Moscardini et al. 1998), whereas the inferred $\epsilon\sim 0$ is more consistent with low $\Omega$ cosmologies. In the models of Moscardini et al. (1998), $\epsilon\sim 0$ for galaxies from $z\sim 0$ to $z\sim 1.5$
 would only be consistent with $\Omega=1$ if there is a rapid increase of the biasing of galaxies relative to the mass distribution over this redshift range, but this possibility is found to be excluded by the low $\omega(\theta)$ amplitude of the Hubble Deep Field and Canada-France Redshift Survey galaxies. 

Enhanced bias is only seen on the $I$-limited HDF at $z>2.4$ (Magliocchetti and Maddox 1998), which is probably beyond the redshifts reached by our survey (Figure 2). However, if the Lyman Break Objects 
subsequently become passively evolving red galaxies at $z\sim 1.5$--2, 
with very red colours of $I-K^{\prime}\sim 5$, they would then 
be much more prominent in a deep $K$-band sample than at $I$ and shorter wavelengths, and the influence of their strong clustering might extend to lower redshifts. We consider the possibility
that this effect is first seen at our fainter limits of 
$K^{\prime}=19.5$--20, where the $\omega(\theta)$ amplitudes exceed the 
$\epsilon=0$ PLE model, becoming larger at the $K=21.5$ limit of Carlberg et al. (1997). 
 
 The clustering measured by Magliocchetti and Maddox (1998) for HDF galaxies with photometric redshifts $2.4<z<2.8$ corresponds, for a low $\Omega$, to a comoving 
$r_0=6.875\pm 2.7 h^{-1}$ Mpc, which  at $z=2.6$ is equivalent to a stable model
with $r_0=16.15\pm 6.4 h^{-1}$ Mpc. The dot-dash line on Figure 7 shows the $\epsilon=0$ PLE model in which the E/S0 and spiral $r_0$ are both increased to this high value above a redshift $z_{i}$.  To fit the high $\omega(\theta)$ amplitudes at $K^{\prime}\geq 19.5$ requires $z_{i}\sim 1.5$ for $K$-limited
surveys, rather than the $z_{i}\simeq 2.4$ directly observed in the $I$-limited HDF. 
 Consistency of this scenario with the much lower $\omega(\theta)$ amplitudes  seen in the deepest $\lambda \sim 0.4$--$0.9 \mu \rm m$ surveys 
would require there to be a very
strong correlation between galaxy colour and cluster environment at 
$z\sim 1.5$--3. This might result from an extended epoch of galaxy formation in
which galaxies formed significantly earlier, and evolved with shorter star-formation timescales, in denser regions of the Universe.

\subsection{The evolution of the galaxy merger rate}

Infante et al. (1996) reported that, for a sample of $R\leq 21.5$ galaxies, the $\omega(\theta)$ at $\theta<6$ arcsec exceeded the inwards extrapolation of the
$\omega(\theta)$ at large separations by 
 about a factor of 1.8. This excess of close pairs was interpreted as 
consisting of physically interacting or merging galaxies, with the numbers suggesting some increase with redshift in the merger/interaction rate. 
In Paper I, there were more than twice
as many $2\leq \theta\leq 3$ arcsec
separation galaxy pairs as expected by chance  for  the section of the data observed with the Redeye camera, but the  excess of pairs above the higher expectation from the fitted $\omega(\theta)$ was of marginal significance, and  no close pair excess was seen on the other half of the data observed with the  Magic camera. 
In the much larger $R$-band survey of Roche and Eales (1998), one of the observed fields showed a significant excess in $\omega(\theta)$ at $2\leq \theta\leq 5$ relative to the amplitude at $\theta>5$ arcsec, again suggesting merger-rate evolution, but  the second field studied did not show a significant difference in the small and large scale $\omega(\theta)$ amplitudes. 

On the five fields studied here, the observed number of close (1.5--5.0 arcsec separation)
pairs of galaxies exceeds, at $\sim 3.3\sigma$ significance, the expectation from the  
fitted $\omega(\theta)$. Similarly, the 
$\omega(\theta)$ at $\theta<3.2$ arcsec shows an excess over the fitted power-law (Figure 6),   
even if $\omega(\theta)$ even for a steeper power-law of $\omega(\theta)\propto \theta^{-1.05}$, corresponding to the steepest 
$\xi(r)$ observed for any type of galaxy (Guzzo et al. 1997).
 
After adding an  estimate of the number of unresolved $\theta \leq 1.5 $ arcsec pairs, we estimate that $11.0\pm 3.4$ per cent
of $K^{\prime}\leq 19.5$ galaxies belong to $\theta<5$ arcsec pairs in excess of the fitted $\omega(\theta)$, if a $\theta^{-0.8}$ power-law is assumed. This is consistent with the estimated pair fractions on the Redeye fields of Paper I at $K<20$ and on field `e' of Roche and Eales (1998) at $R\leq 23$.
  The clear 
detection of a close pair excess  in this relatively high resolution
($\rm FWHM\simeq 1.13$ arcsec) data suggests that the non-detection  on the Magic fields of Paper I and field `f'
of Roche and Eales (1998) was simply the result of
 the poorer resolution ($\rm FWHM\sim 1.8$ arcsec)  of these datasets.

 The fraction of galaxies in close pairs can be interpreted in terms of an evolving merger/interaction rate, parameterized here as  
$R_{m}(z)\propto R_{m0}(1+z)^m$. Following Roche and Eales (1998),
we assume $f_{pair}(z)\propto R_{m}(z)$ (on the basis that the merger timescale is much shorter than the Hubble time) and normalize $f_{pair}$ at $z=0$ using the Carlberg et al. (1994)
estimate that 4.6 per cent of local galaxies are in pairs of projected 
separation $<19$ $h^{-1}$ kpc.
All such pairs will have an angular separation $\theta<5$ arcsec at 
 angular diameter distances $d_A>784$ $h^{-1}$ Mpc, i.e. at $z>0.455$ in our
chosen cosmology. At lower redshifts, assuming  
$\omega(\theta)\propto \theta^{-0.8}$, the fraction of these close-pair galaxies with $\theta<5$ arcsec
will be $({d_A(z)\over 784 h^{-1}})^{1.2}$.

We then model $f_{pair}$ by summing
\begin{equation}
f_{pair}=f_{pair}(z<0.455)+f_{pair}(z>0.455)
\end{equation}
where
\[
f_{pair}(z<0.455)=0.046 {\int_{0}^{0.455}({d_A(z)\over 784 h^{-1}})^{1.2} N(z)(1
+z)^m dz  \over \int_{0}^{0.455}N(z)dz}
\]
\[
f_{pair}(z>0.455)=0.046 {\int_{0.455}^{6}N(z)(1+z)^m dz  \over \int_{0.455}^{6}N
(z)dz}
\]
over the PLE model $N(z)$ for $K^{\prime}\leq 19.5$ (Figure 3).
This predicts $f_{pair}=4.3$ per cent for no evolution ($m=0$), 8.6 per cent
for $m=1$ and 18.2 per cent for $m=2$, with the observed $f_{pair}=11.0\pm 3.4$ per cent
corresponding to $m=1.33_{-0.51}^{+0.36}$. 

This merger rate evolution can be compared with  the $m=2.2\pm 0.5$ estimated by
Infante et al. (1996) from 
$R<21.5$ galaxies,  the $m=2.01^{+0.52}_{-0.69}$ 
of Roche and Eales (1998) for the $R=21.5$--23.5 range of limits, and the $m=1.2\pm 0.4$ of Neuschaefer
et al. (1997) from deep ($I\leq 25$) HST data. Taken in combination, these 
estimates may hint at a reduction in $m$ with increasing depth. 

Carlberg et al. (1994) predict that $m$ will be sensitive to $\Omega$ but
will also be affected by  any reduction
in the mean mass of galaxies at higher redshift (inevitable if merging is occurring), which would cause the merger rate evolution to depart from a simple 
 $R_{m}(z)\propto (1+z)^m$ form in that deeper surveys would give lower 
estimates of $m$. For example, an $\Omega=0.18$ Universe is predicted to give $m=2.2$ at low redshift, falling to $m=1.2$ at the redshift where the mean galaxy mass is reduced by 32 per cent. Our results and the others
discussed here would be consistent with this scenario, but inconsistent with the
more rapid merger-rate evolution ($m=3.2$--4.5) expected for $\Omega=1$. To accurately determine the form of $R_m(z)$, large and deep surveys with multiple passbands (ideally 
both the near-infra-red and optical) to provide photometric redshift estimates 
will be required. Studies of close pair statistics appear to be critically dependent on data resolution, so any attempt to determine $R_m(z)$ should use either space-based observations or only the sharpest of ground-based images.

\subsection*{Acknowledgements}

  The data
reduction and analysis were carried out at the University of Wales 
Cardiff, using
facilities provided by the UK Starlink project,
funded by the PPARC.
NR acknowledges the support of a PPARC research associateship, CJW a PPARC
studentship.

\end{document}